\pdfoutput=1 
\documentclass[a4paper]{article}

\usepackage[pages=all, color=black, position={current page.south}, placement=bottom, scale=1, opacity=1, vshift=5mm]{background}
\SetBgContents{
	\tt This work is shared under a \href{https://creativecommons.org/licenses/by-sa/4.0/}{CC BY-SA 4.0 license} unless otherwise noted
}      

\usepackage[margin=1in]{geometry} 

\usepackage{amsmath}
\usepackage{amsthm}
\usepackage{amssymb}
\usepackage{booktabs}
\usepackage[utf8]{inputenc}
\usepackage{hyperref}

\usepackage[sort&compress,numbers,square]{natbib}

\title{Reduced Efficiency in the Attentional Network During Distractor Suppression in Mild Cognitive Impairment}

\author{Jatupong~Oboun$^1$$^2$ \and Piyanon~Charoenpoonpanich$^1$$^2$$^3$ \and Anna~Raksapatcharawong$^4$\and Chaipat~Chunharas$^3$\and Itthi~Chatnuntawech$^5$\and Chainarong Amornbunchornvej$^1$\thanks{Corresponding author, email: chainarong.amo@nectec.or.th} \and Sirawaj~Itthipuripat$^2$\thanks{Corresponding author, email: itthipuripat.sirawaj@gmail.com}}

\date{
\begin{small}
	$^1$National Electronics and Computer Technology Center (NECTEC), NSTDA, Pathum Thani, 12120, Thailand \\%
        $^2$Neuroscience Center for Research and Innovation, KMUTT, Bangkok, 10140, Thailand \\
        $^3$Chulalongkorn Cognitive Clinical and Computational Neuroscience, Chulalongkorn University, Bangkok, 10330, Thailand \\
        $^4$Faculty of Medicine, Ramathibodi Hospital, Mahidol University, Bangkok, 10400, Thailand \\
        $^5$National Nanotechnology Center (NANOTEC), NSTDA, Pathum Thani, 12120, Thailand
        \end{small}
}

\begin{document}
\maketitle

\section*{Highlights}
\begin{itemize}
    \item Mild Cognitive Impairment (MCI) is an early stage of cognitive decline that often leads to dementia. Early detection and intervention are critical.
    
    \item This study uses EEG and behavioral data to explore how MCI affects the brain's ability to filter distractions.
    
    \item Significant interactions in Global Efficiency (GE) between cognitive status, task congruency, and saliency provide insights into the neural mechanisms underlying MCI. 
    
    \item These findings enhance our understanding of MCI and could inform early intervention strategies, potentially helping to maintain cognitive function and improve quality of life for individuals with MCI.
\end{itemize}

\begin{abstract}
Mild Cognitive Impairment (MCI) is a critical transitional stage between normal cognitive aging and dementia, making its early detection essential. This study investigates the neural mechanisms of distractor suppression in MCI patients using EEG and behavioral data during an attention-cueing Eriksen flanker task. A cohort of 56 MCIs and 26 healthy controls (HCs) performed tasks with congruent and incongruent stimuli of varying saliency levels. During these tasks, EEG data were analyzed for alpha band coherence's functional connectivity, focusing on Global Efficiency (GE), while Reaction Time (RT) and Hit Rate (HR) were also collected.

Our findings reveal significant interactions between congruency, saliency, and cognitive status on GE, RT, and HR. In HCs, congruent conditions resulted in higher GE (p = 0.0114, multivariate t-distribution correction, MVT), faster RTs ($p<0.0001$, MVT), and higher HRs ($p < 0.0001$, MVT) compared to incongruent conditions. HCs also showed increased GE in salient conditions for incongruent trials (p = 0.0406, MVT). MCIs exhibited benefits from congruent conditions with shorter RTs and higher HRs (both $p < 0.0001$, MVT) compared to incongruent conditions but showed reduced adaptability in GE, with no significant GE differences between conditions. 

These results highlight the potential of alpha band coherence and GE as early markers for cognitive impairment. By integrating GE, RT, and HR, this study provides insights into the interplay between neural efficiency, processing speed, and task accuracy. This approach offers valuable insights into cognitive load management and interference effects, indicating benefits for interventions aimed at improving attentional control and processing speed in MCIs.
\end{abstract}

\section{Introduction}

An estimated 22.7 percent of adults aged 65 and older suffer from Mild Cognitive Impairment (MCI) \citep{Rajan2021MCINumber}, a condition that represents a significant precursor to more severe cognitive disorders such as Alzheimer's disease (AD) \citep{Reisberg2008MCItoAD}. MCI is characterized by cognitive impairments that exceed normal age-related decline but do not substantially interfere with daily life, presenting a crucial window for early intervention \citep{petersen1999MCI,petersen2001MCI}. Detecting and understanding MCI early is pivotal, as timely intervention can delay or prevent progression to dementia, thereby improving the quality of life for millions \citep{petersen2004MCI}.

Despite advances in neuroimaging and electrophysiological techniques, there remains a critical gap in understanding the specific neural mechanisms underlying MCI. One key area of interest is the alpha band (8-12 Hz) coherence, which reflects synchronous neural oscillations and is linked to crucial cognitive functions such as attention, memory, and sensory processing \citep{klimesch2012alpha}. Previous studies have demonstrated that individuals with MCI exhibit altered alpha band coherence and reduced overall functional connectivity, suggesting disruptions in neural communication pathways \citep{fodor2021alphabeta, lejko2020alpha_sysreview}. These neural alterations are believed to contribute to the cognitive deficits observed in MCI, making alpha band coherence a promising biomarker for early detection \citep{zheng2007alpha}. Measures of functional connectivity, such as Global Efficiency (GE), provide insights into the efficiency of information transfer across the brain’s networks, highlighting the neural inefficiencies present in MCI \citep{vandenHeuvel2009GE}.

The Eriksen flanker task, a robust paradigm in cognitive neuroscience, has been widely used to investigate attentional control and conflict processing \citep{eriksen1974}. Participants respond to target stimuli flanked by congruent or incongruent distractors, thus engaging mechanisms of distractor suppression and selective attention. The task's sensitivity to cognitive load and interference, especially with the presence of salient distractors, makes it ideal for examining the subtle cognitive impairments associated with MCI. Recent studies have shown that conflicting sensory information can enhance the neural representations of early selective visuospatial attention, highlighting the importance of this task in cognitive research \citep{sookprao2024IEM}.

In this study, we aim to address the existing gaps by employing an attention-cueing version of the Eriksen flanker task to investigate the neural mechanisms of distractor suppression in MCI patients. By varying the congruency and saliency of stimuli, we seek to elucidate how MCI affects the ability to manage cognitive load and suppress irrelevant information. The integration of EEG data enables us to analyze alpha band coherence and functional connectivity, focusing on GE, while also collecting behavioral measures such as Reaction Time (RT) and Hit Rate (HR).

Our primary hypothesis is that MCI patients will exhibit more pronounced alterations in GE and alpha band coherence compared to healthy controls (HCs), particularly under conditions of high cognitive load and interference. We also hypothesize that these neural measures will correlate with behavioral performance, with MCI patients showing slower RTs and lower HRs. Additionally, we expect that saliency will differentially affect MCI patients, potentially exacerbating their cognitive deficits due to impaired attentional control mechanisms.

By examining the interactions between cognitive load, saliency, and cognitive status, our study aims to provide a comprehensive understanding of the neural and behavioral alterations in MCI. This approach not only enhances our knowledge of the underlying neural mechanisms but also contributes to the development of potential biomarkers for early detection and targeted interventions.

The results of this study are expected to have significant implications for the diagnosis and treatment of MCI. By identifying specific neural markers and behavioral patterns associated with MCI, we can improve screening methods and develop more effective cognitive training programs. Furthermore, understanding how MCI patients process and respond to different types of stimuli can inform the design of interventions aimed at enhancing cognitive control and reducing the impact of cognitive decline.

In conclusion, this study leverages the strengths of EEG and the Eriksen flanker task to explore the complex interplay between neural efficiency, cognitive load management, and attentional control in MCI. By focusing on alpha band coherence and functional connectivity, we aim to uncover critical insights into the neural disruptions that characterize MCI and pave the way for improved diagnostic and therapeutic strategies.

\section{Materials and Methods}

{\label{130442}}

\subsection{Subjects}

A cohort of 56 patients with Mild Cognitive Impairment (MCIs) (mean age 67.9 years, 16 males) and 26 healthy controls (HCs) (mean age 65.8 years, 9 males) participated in the Flanker task experiment conducted between 2019 and 2021 at King Chulalongkorn Memorial Hospital, Bangkok, Thailand, with institutional ethics committee approval (0926/64). All participants had either normal or corrected-to-standard color vision and were free from any neurological or psychiatric conditions. Written consent was obtained from each subject before participation, in line with the guidelines of KMUTT's local institutional review board (IRB). The study adhered to the principles outlined in the Declaration of Helsinki.

\subsection{Stimuli and Tasks}

The experimental paradigm closely follows the design and procedures established in previous research from our lab \citep{sookprao2024IEM}. Participants engaged in an attention-cueing Eriksen flanker task to investigate attentional shifts and response selection processes with congruent and incongruent visual stimuli.

We facilitated stimuli presentation and task execution were facilitated using the Psychophysics Toolbox extension \citep{pelli1997Psychtoolbox,brainard1997Psychtoolbox,kleiner2007Psychtoolbox} for MATLAB 2020 \citep{MATLAB} on a standard personal computer. Participants were positioned 80 cm from an LCD monitor, performing the task in a controlled environment to ensure minimal external distractions and optimal task engagement.

In the flanker task, participants were required to discriminate the shape of a target stimulus, presented among distractors within a circular array on the screen. The task involved varying levels of congruency and saliency, with the target's shape (diamond or hourglass) and the distractors' configuration (congruent or incongruent) being pseudo-randomly determined for each trial. Additionally, trials varied in terms of color saliency, with some trials featuring a color singleton designed to capture attention.

The task consisted of trials divided into blocks, with feedback on performance accuracy and speed. The comprehensive session lasted approximately 2.5 hours, including breaks and setup times, with participants compensated for their participation.

For a comprehensive understanding of the task parameters, trial types, and specific experimental conditions utilized in this study, we refer readers to \cite{sookprao2024IEM}. Building upon the established experimental framework, our investigation zeroes in on the differential impact of saliency on subjects by methodically comparing trials that feature salient stimuli against those that do not. This comparative approach aims to unravel the distinct cognitive and neural dynamics elicited by salient cues, thereby contributing to our understanding of how saliency modulates attentional processes and cognitive performance.

\begin{figure}[h!]
\begin{center}
\includegraphics[width=0.90\columnwidth]{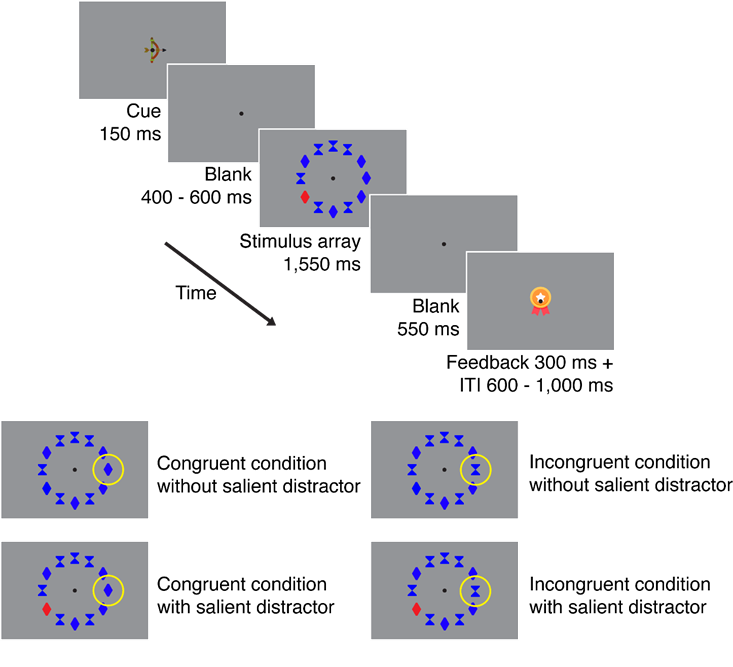}
\caption{
An Adapted Attention-Cueing Flanker Task
(Top) Trial sequences initiated with a fixation-point arrow cue indicating the target location among 12 options, followed by a red or blue circular array of 12 stimuli. Participants identified the target shape (diamond or hourglass) at the cued location, with performance feedback provided post-trial.
(Bottom) Trials featured targets with either congruent or incongruent flankers, based on shape similarity. Half of the trials included a color-distinct salient distractor, while the other half had uniformly colored stimuli.}
\label{fig:Flanker_Diagram}
\end{center}
\end{figure}

\subsection{EEG recording and preprocessing}

In this investigation, EEG data acquisition and preprocessing were conducted following protocols established in prior studies conducted by our lab~\cite{phangwiwat2024sustained}. Continuous EEG signals were captured using a 64-channel ActiveTwo system, complemented by eight supplementary electrodes (Biosemi Instrumentation), at a sampling rate of 512 Hz. This setup encompassed electrodes positioned at the left and right mastoids for referencing, adjacent to the left and right eyes' outer canthi to monitor horizontal eye movements, and above and below both eyes for tracking blinks and vertical eye movements. Initial online referencing was done to the CMS-DRL electrode, with a standard practice of maintaining data offsets across all channels below 20 mV.

For data preprocessing, we employed EEGLab v2020.0 \citep{delorme2004eeglab} in conjunction with customized MATLAB scripts, mirroring the methodology described in~\cite{phangwiwat2024sustained}. The EEG data were first re-referenced to the mean of the left and right mastoid electrodes and filtered using 0.25-Hz high-pass and 55-Hz low-pass Butterworth filters (third order). Epochs were then extracted from the continuous data, ranging from 1,850 ms pre-cue to 5,150 ms post-cue, followed by the application of Independent Component Analysis (ICA) to identify and remove artifacts related to eye movements, muscle activity, and faulty channels \citep{makeig1996ICA}. Artifactual epochs, identified through manual inspection and threshold-based criteria, were discarded, leading to an average exclusion of 23.90 ± 7.89\% SD of trials across participants, as detailed in~\cite{phangwiwat2024sustained}.

\subsection{EEG Functional Connectivity Network (FCN)}

Functional connectivity is a pivotal concept in neuroscience, encapsulating the statistical dependencies that exist among neurophysiological events occurring in spatially distinct brain regions \citep{friston2011functionalconnectivity,Cao2021Functionalconnectivity}. In this study, coherence serves as a fundamental metric for quantifying functional connectivity, representing the degree of phase consistency or linear correlation between pairs of EEG signals across different brain areas. The employment of coherence in our analysis is motivated by its ability to capture the dynamic interactions and integrative processes that underlie cognitive functioning. It provides a quantifiable measure of how different regions of the brain synchronize their activity, thereby facilitating coordinated responses to cognitive demands \citep{bowyer2016coherence,murias2007coherence,srinivasan2007coherence}. This synchronization, or coherence, is especially pertinent within the alpha frequency band, which is associated with attention, relaxation, and various cognitive processes \citep{Jensen2002alpha,lejko2020alpha_sysreview}. By analyzing coherence within the alpha band, we aim to uncover the intricate network dynamics that support cognitive operations and to identify potential alterations.

\begin{figure}[h!]
\begin{center}
\includegraphics[width=0.90\columnwidth]{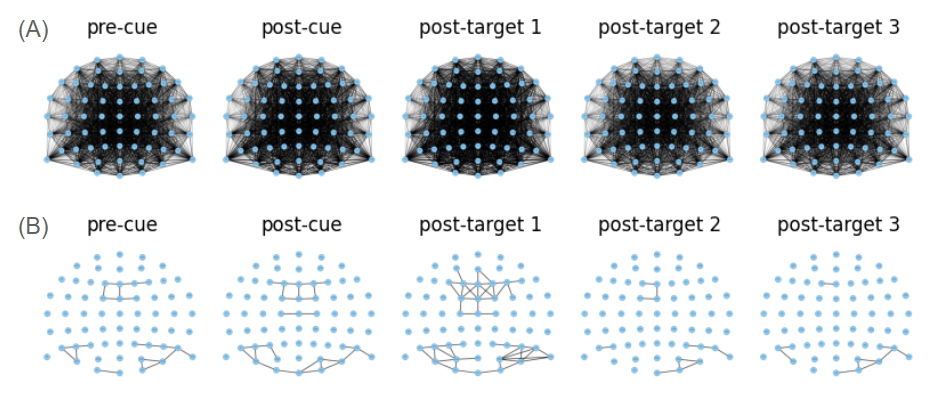}
\caption{Dynamic changes of a trial in one of our subject's FCN during five time periods: pre-cue (-500ms to 0ms window, before cue appear), post-cue (0ms to 500ms window, after cue appear), post-target 1 (0 to 500ms window, after target appear), post-target 2 (500ms to 1000ms window, after target appear), and post-target 3 (1000ms to 1500ms window, after target appear). (A) shows the full connectivity matrix before thresholding, while (B) depicts the network after applying a threshold to emphasize stronger connections.}
\label{fig:FCN}
\end{center}
\end{figure}

The construction of the EEG Functional Connectivity Network (FCN) is centered around assessing the synchrony between all possible pairwise combinations of EEG channels within the alpha frequency band (8 to 12 Hz), using the \texttt{spectral\_connectivity\_epochs} function from the \texttt{mne\_connectivity} library in Python \citep{MNE}. This function computes spectral connectivity measures, including coherence, over trials. Coherence is determined using multitaper methods that are part of the function's capabilities. High coherence values in the alpha band suggest a robust functional connection, indicative of efficient inter-regional communication and integration, which is essential for cognitive processing. The resulting coherence networks are visualized using an implementation of the EEGRAPH library in Python with specific customization \citep{EEGraph}, as shown in Figure \ref{fig:FCN}.

To illustrate the dynamic nature of FCNs, we analyzed connectivity during five critical time windows associated with distinct cognitive events: pre-cue, post-cue, post-target 1, post-target 2, and post-target 3, as illustrated in Figure~\ref{fig:FCN}. This approach allows us to capture the transient reconfigurations of the network that correspond to specific task demands, highlighting the brain's adaptive mechanisms.

\subsubsection{Network Thresholding}

In our analysis, we used a threshold of three standard deviations above the median connectivity for the functional connectivity networks (FCNs). We pooled FCNs from all subjects, time windows, and task conditions. Then we applied this threshold within each frequency band separately. This methodological choice was guided by the observation that FCNs in the lower frequency band generally exhibit stronger connections compared to those in the higher frequency band \citep{cohen2014book}.

\subsection{Global Efficiency (GE)}

Global Efficiency (GE) is a prominent metric in neuroscience used to quantify the efficiency of information transfer within a network. It measures the average inverse shortest path length between all pairs of nodes, reflecting the network's ability to facilitate rapid and efficient communication. This metric has become well-established for investigating the brain's functional integration, especially under varying cognitive states \citep{ismail2020graphanalysis, yu2016global_eff_example, achard2007globalefficiency}.

In our analysis, GE was applied specifically to the \texttt{post-target 1} phase of our thresholded FCNs, corresponding to the 0 to 500 ms window after target appearance. This time window was strategically chosen because it encompasses the immediate responses to the salient and congruent conditions of the attention task, which are crucial for evaluating cognitive processing and detecting potential group differences in brain dynamics. The hypothesis is that during this critical period, cognitive demands are heightened as participants respond to these specific task conditions, thereby making it an optimal point for observing key neural mechanisms and interactions.

The calculation of GE follows:
\[
GE = \frac{1}{N(N-1)} \sum_{i \neq j \in N} \frac{1}{d(i,j)}
\]
where \( N \) represents the number of nodes in the network, and \( d(i, j) \) is the shortest path length between nodes \( i \) and \( j \). The computation of GE was implemented using the NetworkX library in Python, leveraging its robust tools for graph analysis \citep{hagberg2008networkx}.

This metric is instrumental in identifying variations in cognitive performance, providing insights into how different conditions and group dynamics influence neural efficiency within our study's framework.

\section{Statistical Analysis}
\label{sec:StatAna}

All statistical analyses were conducted using R (version 4.0.5) \citep{R}. Linear mixed-effects models (LMMs) were used to assess the effects of Group (HC, MCI), Congruent (CON, INC), and Salient (NS, SA) on three dependent variables: Global Efficiency (GE), Reaction Time (RT), and Hit Rate (HR). Each model included random intercepts for subjects to account for repeated measures. The LMMs were estimated using the \texttt{lmerTest} Package~\citep{lmerTest}.

An LMM was fitted for each dependent variable (GE, RT, HR) with Group, Congruent, and Salient as fixed effects, including all two-way and three-way interactions:

\begin{small}
\begin{verbatim}
lmerTest::lmer(Var ~ Group * Congruent * Salient + (1 | subjectID), data = df)
\end{verbatim}
\end{small}

The significance of fixed effects was evaluated using Type III ANOVA with Satterthwaite's method for approximate degrees of freedom. Estimated marginal means and pairwise comparisons were calculated using the \texttt{emmeans} package \citep{emmeans}, with multiple testing corrections applied using the multivariate t-distribution (MVT) method. Simple effects analyses were conducted to explore significant interactions.

Given a significant three-way interaction in GE, separate two-way LMMs were conducted for each group (Healthy and MCI) to investigate the interaction between Congruent and Salient conditions:

\begin{small}
\begin{verbatim}
lmerTest::lmer(GE ~ Congruent * Salient + (1 | subjectID), data = healthy_data)
lmerTest::lmer(GE ~ Congruent * Salient + (1 | subjectID), data = mci_data)
\end{verbatim}
\end{small}

The significance of fixed effects in the two-way models was evaluated using Type III ANOVA with Satterthwaite's method for approximate degrees of freedom. Estimated marginal means and pairwise comparisons were calculated using the \texttt{emmeans} package \citep{emmeans}, with multiple testing corrections applied using the multivariate t-distribution (MVT) method. Simple effects analyses were conducted to explore significant interactions within each group.
\section{Results}

\subsection{Global Efficiency Results}

\subsubsection{Three-Way Linear Mixed-Effects Model}
The linear mixed-effects model revealed a significant three-way interaction between Group, Congruent, and Salient (\( F(1, 240) = 9.8769 \), \( p = 0.0019 \), Satterthwaite's method). The main effects and other interactions were not significant (Table \ref{tab:GE_LMM}).

\begin{table}[h!]
\centering
\begin{tabular}{lrrrrrr}
\toprule
Effect & Sum Sq & Mean Sq & NumDF & DenDF & \( F \)-value & \( p \)-value \\
\midrule
Group & 2.1800e-06 & 2.1800e-06 & 1 & 80 & 0.1365 & 0.7128 \\
Congruent & 2.4797e-05 & 2.4797e-05 & 1 & 240 & 1.5524 & 0.2140 \\
Salient & 3.7000e-08 & 3.7000e-08 & 1 & 240 & 0.0023 & 0.9617 \\
Group:Congruent & 1.8790e-06 & 1.8790e-06 & 1 & 240 & 0.1176 & 0.7319 \\
Group:Salient & 3.8360e-06 & 3.8360e-06 & 1 & 240 & 0.2401 & 0.6246 \\
Congruent:Salient & 2.4047e-05 & 2.4047e-05 & 1 & 240 & 1.5054 & 0.2210 \\
\textbf{Group:Congruent:Salient} & \textbf{1.5777e-04} & \textbf{1.5777e-04} & \textbf{1} & \textbf{240} & \textbf{9.8769} & \textbf{0.0019} \\
\bottomrule
\end{tabular}
\caption{Results of the linear mixed-effects model with Satterthwaite's method.}
\label{tab:GE_LMM}
\end{table}

\subsubsection{Two-Way Linear Mixed-Effects Models for Each Group}

\paragraph{Healthy Controls (HCs)}
The two-way interaction between Congruent and Salient was significant (\( F(1, 75) = 4.3622 \), \( p = 0.04014 \), Satterthwaite's method). This indicates that the impact of Saliency on Global Efficiency depends on whether the task is Congruent or Incongruent in the Healthy group.

\begin{table}[h!]
\centering
\begin{tabular}{lrrrrrr}
\toprule
Effect & Sum Sq & Mean Sq & NumDF & DenDF & \( F \)-value & \( p \)-value \\
\midrule
Congruent & 1.4763e-05 & 1.4763e-05 & 1 & 75 & 0.5768 & 0.44996 \\
Salient & 1.1420e-05 & 1.1420e-05 & 1 & 75 & 0.0446 & 0.83329 \\
\textbf{Congruent:Salient} & \textbf{1.1165e-04} & \textbf{1.1165e-04} & \textbf{1} & \textbf{75} & \textbf{4.3622} & \textbf{0.04014} \\
\bottomrule
\end{tabular}
\caption{Results of the two-way linear mixed-effects model for the Healthy group.}
\label{tab:HC_2way}
\end{table}

\paragraph{Mild Cognitive Impairment patients(MCIs)}
The two-way interaction between Congruent and Salient was significant (\( F(1, 165) = 3.9850 \), \( p = 0.04755 \), Satterthwaite's method  ). This indicates that the impact of Saliency on Global Efficiency depends on whether the task is Congruent or Incongruent in the MCI group.

\begin{table}[h!]
\centering
\begin{tabular}{lrrrrrr}
\toprule
Effect & Sum Sq & Mean Sq & NumDF & DenDF & \( F \)-value & \( p \)-value \\
\midrule
Congruent & 1.0269e-05 & 1.0269e-05 & 1 & 165 & 0.8852 & 0.34815 \\
Salient & 3.6480e-06 & 3.6480e-06 & 1 & 165 & 0.3144 & 0.57573 \\
\textbf{Congruent:Salient} & \textbf{4.6226e-05} & \textbf{4.6226e-05} & \textbf{1} & \textbf{165} & \textbf{3.9850} & \textbf{0.04755} \\
\bottomrule
\end{tabular}
\caption{Results of the two-way linear mixed-effects model for the MCI group.}
\label{tab:MCI_2way}
\end{table}

\begin{figure}[h!]
    \centering
    \includegraphics[width=0.8\textwidth]{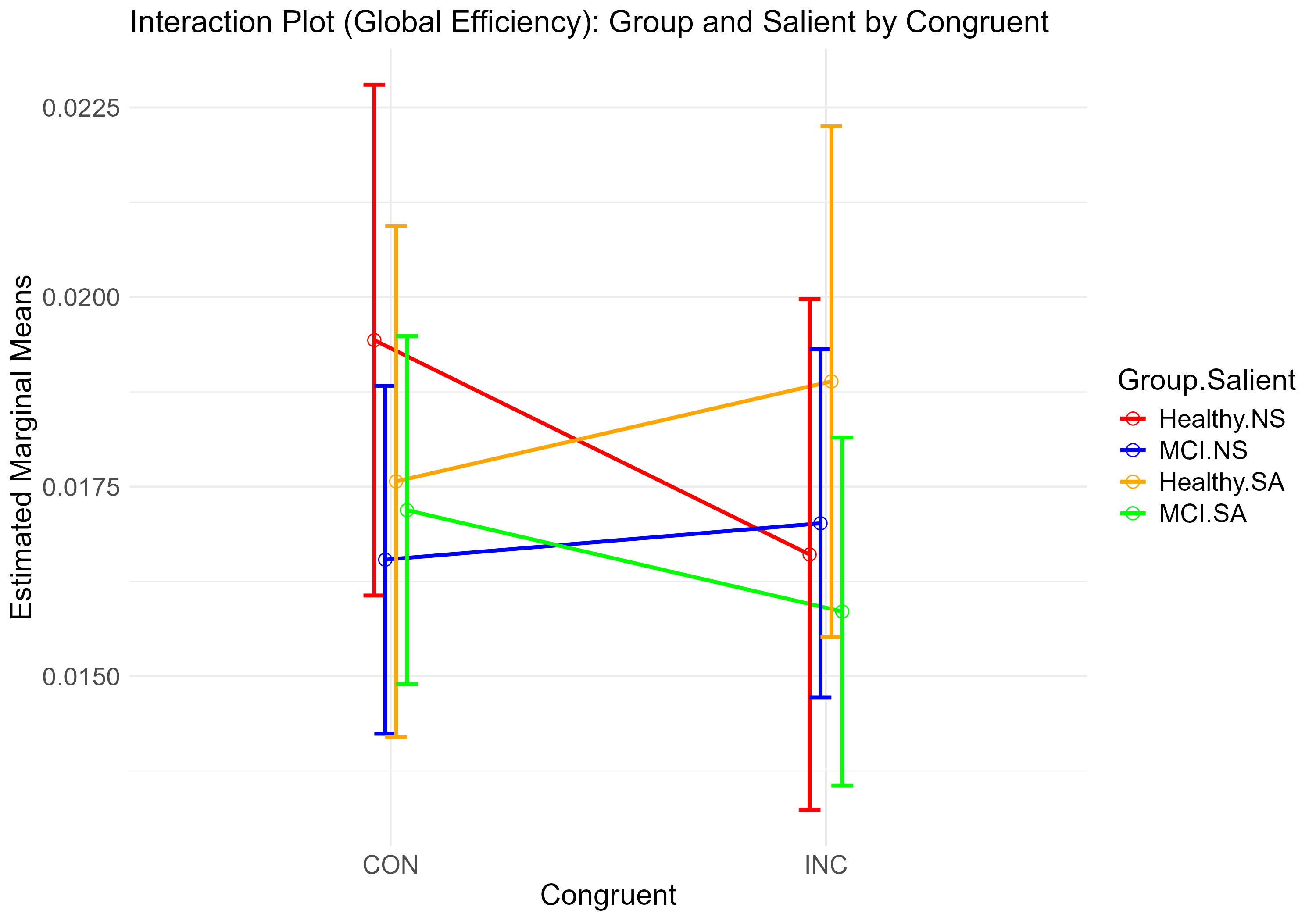}
    \caption{Interaction plot for Global Efficiency: Congruency (CON, INC) and Saliency (NS, SA) effects for each Group (HCs, MCIs). In the HCs, GE for CON is significantly higher than for INC in the NS condition (\( p = 0.0114 \)). In the SA condition, GE for INC is higher than for CON, but this difference is not significant. In the HCs, the SA condition shows a significant increase in GE for INC compared to NS (\( p = 0.0406 \)). The CON condition shows no significant difference between SA and NS. In the MCIs, there is no significant variation in GE across Congruent and Salient conditions, indicating that saliency and congruency have less impact on this group. Overall, the Healthy group's performance is more sensitive to changes in saliency and congruency compared to the MCI group.}
    \label{fig:interaction_summary}
\end{figure}

The interaction between Congruent and Salient conditions for each group is visualized in Figure \ref{fig:interaction_summary}.

\subsubsection{Simple Effects Analyses}

\paragraph{Group}
None of the comparisons for Group within Congruent and Salient conditions were significant (Table \ref{tab:GE_Group}).

\begin{table}[h!]
\centering
\begin{tabular}{lrrrrrr}
\toprule
Salient & Congruent & Contrast      & Estimate & SE       & \( t \)-ratio & \( p \)-value \\
\midrule
NS      & CON       & HCs vs MCIs & 0.001448 & 0.00204  & 0.711  & 0.4793 \\
SA      & CON       & HCs vs MCIs & 0.000190 & 0.00204  & 0.093  & 0.9261 \\
NS      & INC       & HCs vs MCIs & -0.000205 & 0.00204 & -0.101 & 0.9199 \\
SA      & INC       & HCs vs MCIs & 0.001517  & 0.00204 & 0.745  & 0.4584 \\
\bottomrule
\end{tabular}
\caption{Simple effects for Group within each Congruent and Salient condition.}
\label{tab:GE_Group}
\end{table}

\paragraph{Salient}
In the HCs and INC condition, the difference between NS and SA was significant (\( p = 0.0406 \), multivariate t-distribution corrected), with visualization shows that the NS condition had a significantly lower GE than the SA condition (Table \ref{tab:GE_Salient}). No other comparisons for Salient within Group and Congruent conditions were significant.

\begin{table}[h!]
\centering
\begin{tabular}{lrrrrrr}
\toprule
Congruent & Group   & Contrast & Estimate & SE       & \( t \)-ratio & \( p \)-value \\
\midrule
CON       & HCs & NS vs SA & 0.000931  & 0.000554 & 1.680  & 0.0942 \\
\textbf{INC}       & \textbf{HCs} & \textbf{NS} vs \textbf{SA} & \textbf{-0.001141} & \textbf{0.000554} & \textbf{-2.059} & \textbf{0.0406} \\
CON       & MCIs     & NS vs SA & -0.000327 & 0.000378 & -0.865 & 0.3879 \\
INC       & MCI s    & NS vs SA & 0.000582  & 0.000378 & 1.541  & 0.1247 \\
\bottomrule
\end{tabular}
\caption{Simple effects for Salient within each Group and Congruent condition.}
\label{tab:GE_Salient}
\end{table}

\paragraph{Congruent}
In the HCs and NS condition, the difference between CON and INC was significant (\( p = 0.0114 \), multivariate t-distribution corrected), with visualization shows that the CON condition had a significantly higher GE than the INC condition (Table \ref{tab:GE_Congruent}). No other comparisons for Congruent within Group and Salient conditions were significant.

\begin{table}[h!]
\centering
\begin{tabular}{lrrrrrr}
\toprule
Salient & Group & Contrast & Estimate & SE & \( t \)-ratio & \( p \)-value \\
\midrule
\textbf{NS} & \textbf{HCs} & \textbf{CON} vs \textbf{INC} & \textbf{-0.001413} & \textbf{0.000554} & \textbf{-2.549} & \textbf{0.0114} \\
SA & HCs & CON vs INC & 0.000659 & 0.000554 & 1.190 & 0.2353 \\
NS & MCIs & CON vs INC & 0.000240 & 0.000378 & 0.636 & 0.5254 \\
SA & MCIs & CON vs INC & -0.000668 & 0.000378 & -1.770 & 0.0780 \\
\bottomrule
\end{tabular}
\caption{Simple effects for Congruent within each Group and Salient condition.}
\label{tab:GE_Congruent}
\end{table}

\subsection{Reaction Time Results}

\subsubsection*{Linear Mixed-Effects Model}
The linear mixed-effects model revealed significant effects for Group (\( F(1, 80) = 5.4936 \), \( p = 0.021568 \), Satterthwaite's method), Congruent (\( F(1, 240) = 371.6789 \), \( p < 2.2 \times 10^{-16} \), Satterthwaite's method), and the interaction between Group and Congruent (\( F(1, 240) = 7.6858 \), \( p = 0.006002 \), Satterthwaite's method). The main effect of Salient and other interactions were not significant (Table \ref{tab:RT_LMM}). 

\begin{table}[h]
\centering
\begin{tabular}{lrrrrrr}
\toprule
Effect & Sum Sq & Mean Sq & NumDF & DenDF & \( F \)-value & \( p \)-value \\
\midrule
\textbf{Group} & \textbf{0.004130} & \textbf{0.004130} & \textbf{1} & \textbf{80} & \textbf{5.4936} & \textbf{0.021568} \\
Salient & 0.000022 & 0.000022 & 1 & 240 & 0.0287 & 0.865537 \\
\textbf{Congruent} & \textbf{0.279424} & \textbf{0.279424} & \textbf{1} & \textbf{240} & \textbf{371.6789} & \textbf{$<$ 2.2e-16} \\
Group:Salient & 0.000075 & 0.000075 & 1 & 240 & 0.1003 & 0.751773 \\
\textbf{Group:Congruent} & \textbf{0.005778} & \textbf{0.005778} & \textbf{1} & \textbf{240} & \textbf{7.6858} & \textbf{0.006002} \\
Salient:Congruent & 0.000058 & 0.000058 & 1 & 240 & 0.0768 & 0.781961 \\
Group:Salient:Congruent & 0.000046 & 0.000046 & 1 & 240 & 0.0617 & 0.803977 \\
\bottomrule
\end{tabular}
\caption{Results of the linear mixed-effects model with Satterthwaite's method.}
\label{tab:RT_LMM}
\end{table}

\subsubsection{Interaction Between Congruent and Group}
\begin{figure}[h!]
    \centering
    \includegraphics[width=0.8\textwidth]{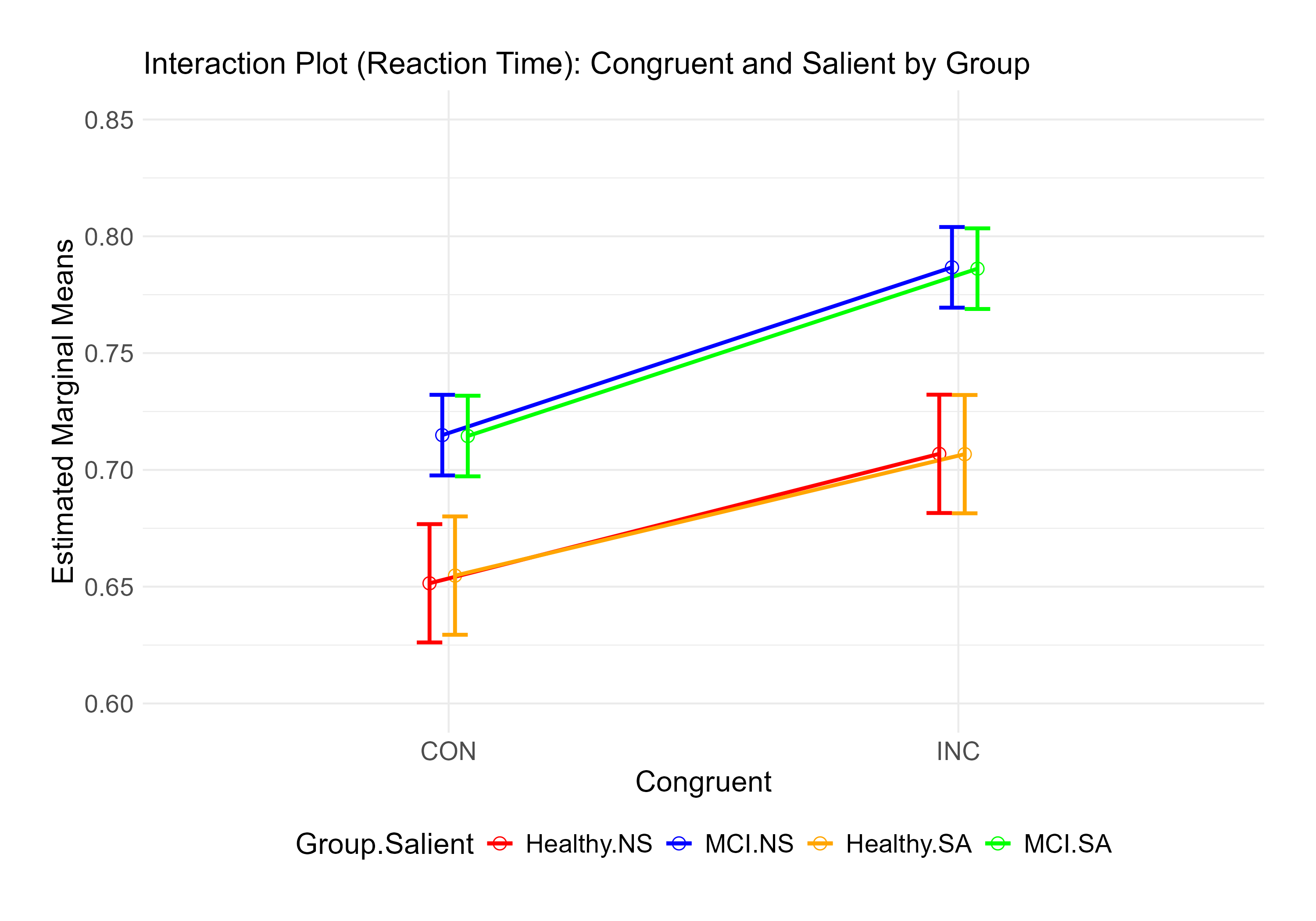}
    \caption{Interaction plot showing the interaction between Congruent (CON, INC) and Group (HCs, MCIs) conditions for Reaction Time (RT). Significant main effects for Group and Congruent were found, as well as a significant interaction between Group and Congruent.}
    \label{fig:RT_Con_Group_interaction}
\end{figure}

The interaction between Congruent (CON, INC) and Group (HCs, MCIs) is shown in Figure \ref{fig:RT_Con_Group_interaction}. In the Healthy group, RT for the CON condition is significantly lower than for the INC condition (\( p < 2.2e-16 \), multivariate t-distribution corrected). This trend is consistent across both Salient (NS, SA) conditions. For the MCI group, there is a significant difference between CON and INC (\( p = 0.006 \), multivariate t-distribution corrected), with CON having a lower RT than INC. For the CON condition, the difference between the HCs and MCI groups was significant (\( p = 0.0103 \), multivariate t-distribution corrected), indicating that the Healthy group had a significantly different RT compared to the MCI group. Similarly, for the INC condition, the difference between the HCs and MCIs was also significant (\( p = 0.0453 \), multivariate t-distribution corrected).

\subsubsection{Simple Effects Analyses}

\paragraph{Congruent}
In both the HCs and MCIs, the difference between CON and INC was significant (\( p < 0.0001 \), multivariate t-distribution corrected) (Table \ref{tab:RT_Congruent}). This indicates that the CON condition had a significantly different RT compared to the INC condition for both groups.

\begin{table}[h!]
\centering
\begin{tabular}{lrrrrr}
\toprule
Group & Contrast & Estimate & SE & \( t \)-ratio & \( p \)-value \\
\midrule
HCs & CON vs INC & 0.0269 & 0.00269 & 9.987 & \textbf{$<$ 0.0001} \\
MCIs & CON vs INC & 0.0359 & 0.00183 & 19.581 & \textbf{$<$ 0.0001} \\
\bottomrule
\end{tabular}
\caption{Simple effects for Congruent within each Group condition. Results are averaged over the levels of Salience.}
\label{tab:RT_Congruent}
\end{table}

\paragraph{Group}
For the CON condition, the difference between the HCs and MCIs was significant (\( p = 0.0103 \), multivariate t-distribution corrected) (Table \ref{tab:RT_Group}), indicating that the HCs had a significantly different RT compared to the MCI group. Similarly, for the INC condition, the difference between the HCs and MCIs was also significant (\( p = 0.0453 \), multivariate t-distribution corrected).

\begin{table}[h!]
\centering
\begin{tabular}{lrrrrr}
\toprule
Congruent & Contrast & Estimate & SE & \( t \)-ratio & \( p \)-value \\
\midrule
CON & HCs vs MCIs & -0.0308 & 0.0152 & -2.033 & \textbf{0.0453} \\
INC & HCs vs MCIs & -0.0398 & 0.0152 & -2.628 & \textbf{0.0103} \\
\bottomrule
\end{tabular}
\caption{Simple effects for Group within each Congruent condition. Results are averaged over the levels of Salience.}
\label{tab:RT_Group}
\end{table}

\subsection{Hit Rate Results}

\subsubsection*{Linear Mixed-Effects Model}
The linear mixed-effects model revealed a significant main effect for Congruent (\( F(1, 240) = 677.6975 \), \( p < 2 \times 10^{-16} \), Satterthwaite's method). The main effects of Group, Salient, and other interactions were not significant (Table \ref{tab:HR_LMM}). The main effect plot of Congruent (CON, INC) is shown in Figure \ref{fig:HR_Con_main_effect}.

\begin{table}[h]
\centering
\begin{tabular}{lrrrrrr}
\toprule
Effect & Sum Sq & Mean Sq & NumDF & DenDF & \( F \)-value & \( p \)-value \\
\midrule
Group & 0.00025 & 0.00025 & 1 & 80 & 0.0952 & 0.7585 \\
Salient & 0.00247 & 0.00247 & 1 & 240 & 0.9485 & 0.3311 \\
\textbf{Congruent} & \textbf{1.76671} & \textbf{1.76671} & \textbf{1} & \textbf{240} & \textbf{677.6975} & \textbf{$<$ 2e-16} \\
Group:Salient & 0.00063 & 0.00063 & 1 & 240 & 0.2421 & 0.6231 \\
Group:Congruent & 0.00090 & 0.00090 & 1 & 240 & 0.3459 & 0.5570 \\
Salient:Congruent & 0.00247 & 0.00247 & 1 & 240 & 0.9459 & 0.3317 \\
Group:Salient:Congruent & 0.00001 & 0.00001 & 1 & 240 & 0.0049 & 0.9440 \\
\bottomrule
\end{tabular}
\caption{Results of the linear mixed-effects model with Satterthwaite's method for Hit Rate.}
\label{tab:HR_LMM}
\end{table}

\subsubsection{Main Effect of Congruent}
\begin{figure}[h!]
    \centering
    \includegraphics[width=0.8\textwidth]{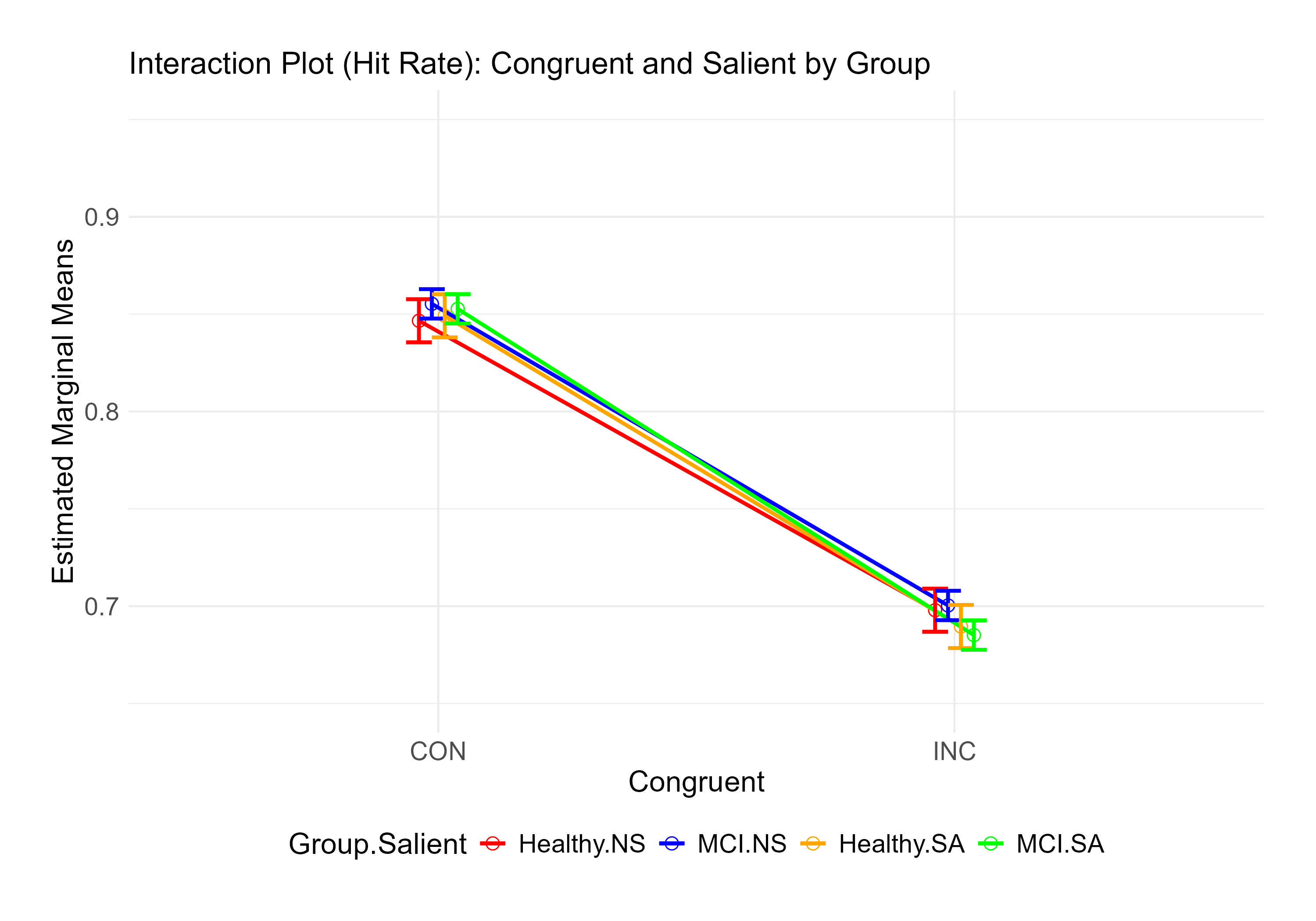}
    \caption{Main effect of Congruent (CON, INC) on Hit Rate (HR) averaged over Group (HCs, MCIs) and Salient (NS, SA) conditions. HR is significantly higher for the CON condition compared to the INC condition (\( p < 2e-16 \), Satterthwaite's method).}
    \label{fig:HR_Con_main_effect}
\end{figure}

The main effect of Congruent (CON, INC) on Hit Rate (HR) is shown in Figure \ref{fig:HR_Con_main_effect}. HR is significantly higher for the CON condition compared to the INC condition (\( p < 2e-16 \), Satterthwaite's method). This effect was consistent across both Group (Healthy, MCI) and Salient (NS, SA) conditions, indicating that congruency significantly impacts HR regardless of the other variables.

\section{Discussion}

\subsection{Overview}

This study investigated the neural mechanisms of distractor suppression in Mild Cognitive Impairment patients (MCIs) using EEG and behavioral data, specifically examining Global Efficiency (GE), Reaction Time (RT), and Hit Rate (HR) during an attention-cueing Eriksen flanker task. The results reveal significant interactions and main effects that provide insights into how MCIs affects cognitive processing and attentional mechanisms. 

\subsection{Global Efficiency}

The analysis of Global Efficiency (GE) on alpha band coherence's functional connectivity network during stimulus period revealed a significant three-way interaction between Group, Congruent, and Salient conditions (Table \ref{tab:GE_LMM}). 

In healthy controls (HC), there is a significant two-way interaction between Congruenct and Salient conditions(\( F(1, 75) = 4.3622 \), \( p = 0.04014 \), Satterthwaite's method) (Table \ref{tab:HC_2way}) which further suggest the effect of congruency and saliency on GE. A deeper analysis of simple effects showed that GE was significantly higher in the congruent (CON) condition compared to the incongruent (INC) condition under non-salient (NS) trials (p = 0.0114, multivariate t-distribution corrected). In salient (SA) trials, the difference between CON and INC conditions was smaller and reversed in direction. Additionally, in HC, GE was significantly higher in the SA condition compared to the NS condition for INC trials (p = 0.0406, multivariate t-distribution corrected.) However, this effect was reversed and not statistically significant in the CON trials \ref{fig:interaction_summary}. In MCI group, despite having two-way interaction between Congruent and Salient conditions was significant (\( F(1, 165) = 3.9850 \), \( p = 0.04755 \)), which indicating that the relationship between these factors affects GE. However, simple effects analyses did not reveal significant differences within each condition (Table \ref{tab:GE_Group}). 

Our findings on the alterations in GE and alpha band coherence in MCI align with previous research, which indicates that alpha band activity is crucial for cognitive processes such as attention and working memory. \cite{fodor2021alphabeta} reported that MCI patients show altered alpha and beta band functional connectivity, which is linked to cognitive impairments in tasks requiring sustained attention and cognitive control. Similarly, \cite{zheng2007alpha} examined EEG spectral power and inter-/intra-hemispheric coherence on alpha frequency bands in MCI patients during a working memory task. They found that MCI patients exhibit greater inter-hemispheric coherence than intra-hemispheric coherence when memory demands increase, suggesting a compensatory mechanism to maintain cognitive processing. In addition, a meta-analysis by \cite{lejko2020alpha_sysreview} found that MCI patients show lower alpha power and functional connectivity during tasks of executive functions compared to cognitively healthy older adults. This study supports that alpha band is an early marker of cognitive decline.

These findings can also be linked to the salience network's role in cognitive processing as it can contextualized within the theoretical framework proposed by \cite{menon2010saliency}, which emphasizes the role of the anterior insula (AI) and anterior cingulate cortex (ACC) in the salience network (SN). According to their model, the AI functions as a critical hub for detecting salient events and facilitating network switching between the default mode network (DMN) and the central executive network (CEN) to allocate cognitive resources efficiently. This mechanism is crucial for processing salient stimuli, which can influence attentional control and cognitive flexibility. Our results show higher global efficiency (GE) in congruent conditions under non-salient trials and significant differences in GE between salient and non-salient conditions during incongruent trials. However, the interaction in mild cognitive impairment (MCI) patients is not as strong as in healthy controls (HC). This supports the idea that the SN is important for modulating attentional processes and might be a potential biomarker for detecting MCI.

Together, these studies support the notion that alpha band coherence and GE are valuable tools for investigating cognitive function changes in MCI patients compared to HC. These findings highlight the potential of GE as an early marker of cognitive impairment. In our study, HC showed significant changes in certain conditions, involving congruency and saliency, while MCI showed much less. These GE changes suggest that HC can dynamically adjust their neural networks to efficiently process varying levels of cognitive demand, indicating the presence of compensatory mechanisms. In contrast, MCI patients did not show such significant differences in GE, suggesting an impaired ability to adaptively reconfigure their neural networks in response to cognitive demands. This inability to modulate GE in response to different types of stimuli may reflect a breakdown in the compensatory mechanisms that are still functional in HC. Therefore, the differential GE responses between HC and MCI patients highlight both the presence of compensatory mechanisms in HC and their potential decline in MCI, serving as early markers of cognitive impairment.

\subsection{Reaction Time}

The analysis of Reaction Time (RT) revealed significant effects for Group (\( F(1, 80) = 5.4936 \), \( p = 0.021568 \), Satterthwaite's method), Congruent (\( F(1, 240) = 371.6789 \), \( p < 2.2 \times 10^{-16} \), Satterthwaite's method), and the interaction between Group and Congruent conditions (\( F(1, 240) = 7.6858 \), \( p = 0.006002 \), Satterthwaite's method) (Table \ref{tab:RT_LMM}). The simple effects analysis revealed that in both the Healthy Controls (HC) and Mild Cognitive Impairment (MCI) groups, RTs were significantly faster in the congruent (CON) condition compared to the incongruent (INC) condition for both groups (p<0.0001, multivariate t-distribution corrected) (Table \ref{tab:RT_Congruent}). Additionally, the difference between the Healthy and MCI groups was significant for both the CON (p=0.0453, multivariate t-distribution corrected) and INC (p=0.0103, multivariate t-distribution corrected) conditions (Table \ref{tab:RT_Group}). These results indicate that both the congruency of stimuli and the cognitive status of participants significantly influence RT.

In the HCs, participants exhibited significantly faster RTs in the CON condition compared to the INC condition. This pattern aligns with existing literature that suggests congruent stimuli reduce cognitive load and enhance processing speed by minimizing interference from distractors \citep{eriksen1974}. The significant difference in RT between the CON and INC conditions highlights the efficiency of attentional control mechanisms in healthy individuals when processing congruent stimuli. Specifically, the simple effects analysis revealed a significant difference between CON and INC conditions in the HCs (p<0.0001, multivariate t-distribution corrected) (Table \ref{tab:RT_Congruent}).

Similarly, the Mild Cognitive Impairment patients (MCIs) also demonstrated a significant difference between the CON and INC conditions, with faster RTs in the CON condition. However, the overall RTs were slower compared to the HCs, reflecting the cognitive impairments associated with MCIs. This finding is consistent with previous research showing that MCI patients experience a general slowing of cognitive processing, yet still benefit from congruent conditions \citep{belleville2007working,belleville2011working}. The simple effects analysis for the MCIs also showed a significant difference between the CON and INC conditions (p<0.0001, multivariate t-distribution corrected) (Table \ref{tab:RT_Congruent}).

The significant interaction between Group and Congruent conditions suggests that while both HCs and MCIs benefit from congruent stimuli, HC participants generally exhibit faster RTs. However, it's crucial to note that the main effects for Group and Congruent were also significant. This indicates that both the cognitive status of participants and the congruency of stimuli independently affect RT, in addition to their interaction effect. While the interaction highlights a difference in the extent of the benefit from congruency between groups, the separate main effects underscore that each factor alone also has a substantial impact on RT.

Moreover, the non-significant main effect of salient conditions indicates that the saliency of the stimuli did not significantly affect reaction time (RT). This could be due to the fact that both HC and MCI participants were equally capable of leveraging the saliency cues to aid their performance by effectively managing and filtering out these distractions. This is supported by studies showing that while Alzheimer's disease (AD) patients exhibit baseline shifts in median hit RT for simple feature search tasks, indicating generalized slowing, they show disproportionate increases in RT for conjoined feature search tasks as array size increases. \cite{foster1999salientnoef} found that both HC and MCI participants were able to manage distractions in simpler tasks, maintaining their performance despite the presence of salient distractors. These findings suggest that while attentional control and the ability to handle distractions are impaired in more complex tasks, the use of saliency cues to manage distractions in simpler tasks might still be effectively preserved in both HC and MCI populations.

Overall, these findings emphasize the importance of congruency in facilitating faster cognitive processing across both healthy and cognitively impaired populations. The results suggest potential benefits for cognitive interventions targeting attentional control and processing speed in MCI patients. By incorporating tasks that emphasize congruent stimuli, it may be possible to enhance cognitive performance in this population, leveraging preserved cognitive mechanisms to mitigate the impact of cognitive decline.

\subsection{Hit Rate}
The analysis of Hit Rate (HR) revealed a significant main effect for Congruent (CON) versus Incongruent (INC) conditions (\( F(1, 240) = 677.6975 \), \( p < 2 \times 10^{-16} \)), indicating that participants had a higher HR for CON conditions across both groups (Table \ref{tab:HR_LMM}). This suggests that congruency significantly enhances participants' ability to correctly identify targets.

Both the Healthy Controls (HCs) and the Mild Cognitive Impairment patients (MCIs) showed higher HR in the CON condition compared to the INC condition. This aligns with findings by  \cite{eriksen1974}, who demonstrated that congruent stimuli reduce cognitive load and enhance task performance by minimizing the interference from distractors. The overall HR was similar between HCs and MCIs, indicating that both groups benefit similarly from congruent conditions despite the cognitive impairments associated with MCI.

The lack of significant main effects for Group and Salient conditions, as well as their interactions, suggests that the primary factor influencing HR was the congruency of the stimuli. This is supported by studies like \cite{ciesielski2006nback}, which found that congruent conditions facilitate more efficient neural processing and reduce error rates in both healthy and cognitively impaired populations. The absence of significant group differences further indicates that, despite the cognitive decline in MCI patients, the fundamental mechanism of enhanced performance under congruent conditions remains intact.

Interestingly, the non-significant main effect of Group suggests that, although MCI patients have slower reaction times, their accuracy in detecting targets is not significantly different from healthy controls under congruent conditions. This aligns with findings indicating that, while Alzheimer's disease (AD) patients struggle more with complex tasks involving distractions, both HCs and MCIs can maintain accuracy levels in simpler tasks with fewer distractions. \cite{foster1999salientnoef} demonstrated that HCs and MCIs effectively manage salient distractors in simple feature search tasks, maintaining their accuracy. These findings suggest that while attentional control and the ability to handle distractions are impaired in more complex tasks for MCIs, their performance on simpler tasks, where saliency cues can aid in filtering out distractions, remains comparable to that of healthy controls.

In conclusion, the significant effect of congruency on HR underscores the critical role of congruent stimuli in enhancing target identification accuracy for both HCs and those with MCI. Despite the cognitive decline associated with MCI, the similar HRs between the MCIs and HCs under congruent conditions highlight the preserved ability of MCI patients to benefit from reduced cognitive load and minimized distractor interference. These findings emphasize the importance of incorporating congruent stimuli in cognitive tasks and interventions, suggesting that leveraging congruent conditions could support better cognitive performance and accuracy in target detection for individuals with cognitive impairments.

\subsection{Integrating Global Efficiency, Reaction Time, and Hit Rate}

By examining Global Efficiency (GE), Reaction Time (RT), and Hit Rate (HR) together, we gain a more nuanced understanding of cognitive processing and attentional control, especially in the context of healthy controls (HCs) and those with mild cognitive impairment (MCIs). Each metric offers unique insights, but their combined interpretation provides a richer, more comprehensive picture of cognitive functioning.

\paragraph{Enhanced Understanding Through Integration}

\begin{itemize}
    \item \textbf{Cognitive Processing Dynamics}:
        \begin{itemize}
            \item \textbf{Global Efficiency (GE)} measures the brain's ability to transfer information efficiently across neural networks. Higher GE in HCs during congruent tasks suggests optimal network functionality.
            \item \textbf{Reaction Time (RT)} indicates processing speed. Faster RTs in congruent conditions for both groups, and also overall faster RT in HCs compare to MCIs, highlight the efficiency of attentional control mechanisms.
            \item \textbf{Hit Rate (HR)} reflects task accuracy. Higher HRs in congruent conditions for both groups suggest reduced cognitive load and improved performance.
        \end{itemize}
        \textbf{Integrated Insight}: Combining these measures shows that HCs can dynamically adjust their neural networks (GE) to process information quickly than MCI (faster RT). In contrast, MCI participants show less adaptability in GE, leading to slower RTs. This integrated view underscores the interplay between network efficiency, and speed in cognitive performance.
        
    \item \textbf{Compensatory Mechanisms and Cognitive Resilience}:
        \begin{itemize}
            \item HC show significant adaptability in GE under varying task demands, demonstrating cognitive resilience and effective compensatory mechanisms.
            \item MCI participants, while still benefiting from congruency as having faster RTs and better HRs in CON condision comparing to INC condition. MCIs exhibit reduced GE adaptability, and generally slower RTs, but still similar HRs trend to HC, indicating a decline in cognitive resilience.
        \end{itemize}
        \textbf{Integrated Insight}: The combined measures reveal how cognitive resilience and compensatory mechanisms function in HC, allowing them to maintain high performance under different conditions. For MCI participants, the reduced adaptability and slower response times highlight the need for targeted interventions to support these compensatory mechanisms.
        
    \item \textbf{Impact of Cognitive Load and Interference}:
        \begin{itemize}
            \item Congruent conditions reduce cognitive load, as evidenced by higher GE, faster RTs, and higher HRs in both groups.
            \item Incongruent conditions increase cognitive load, resulting in lower GE, slower RTs, and reduced HRs, with MCI participants being more adversely affected.
        \end{itemize}
        \textbf{Integrated Insight}: This holistic perspective highlights the critical role of cognitive load and interference in modulating cognitive performance. Understanding how GE, RT, and HR interact under different conditions provides insights into the mechanisms underlying cognitive load management and the impact of interference on task performance.
\end{itemize}

\subsection{Exploratory Results in Routing Efficiency}

To probe whether the whole‑brain deficit captured by Global Efficiency (GE)
is accompanied by pathway‑specific changes, we conducted an exploratory
\emph{Routing Efficiency} (RE) analysis.  
RE quantifies the inverse shortest‑path length between the “core”
fronto‑parietal and “extended” temporo‑parietal/occipital subnetworks.
Guided by the GE time‑course, we restricted this exploration to the
\textbf{0–0.5 s window after target onset}—the epoch that showed the
largest GE separation between groups.

\begin{figure}[h!]
    \centering
    \includegraphics[width=0.4\textwidth]{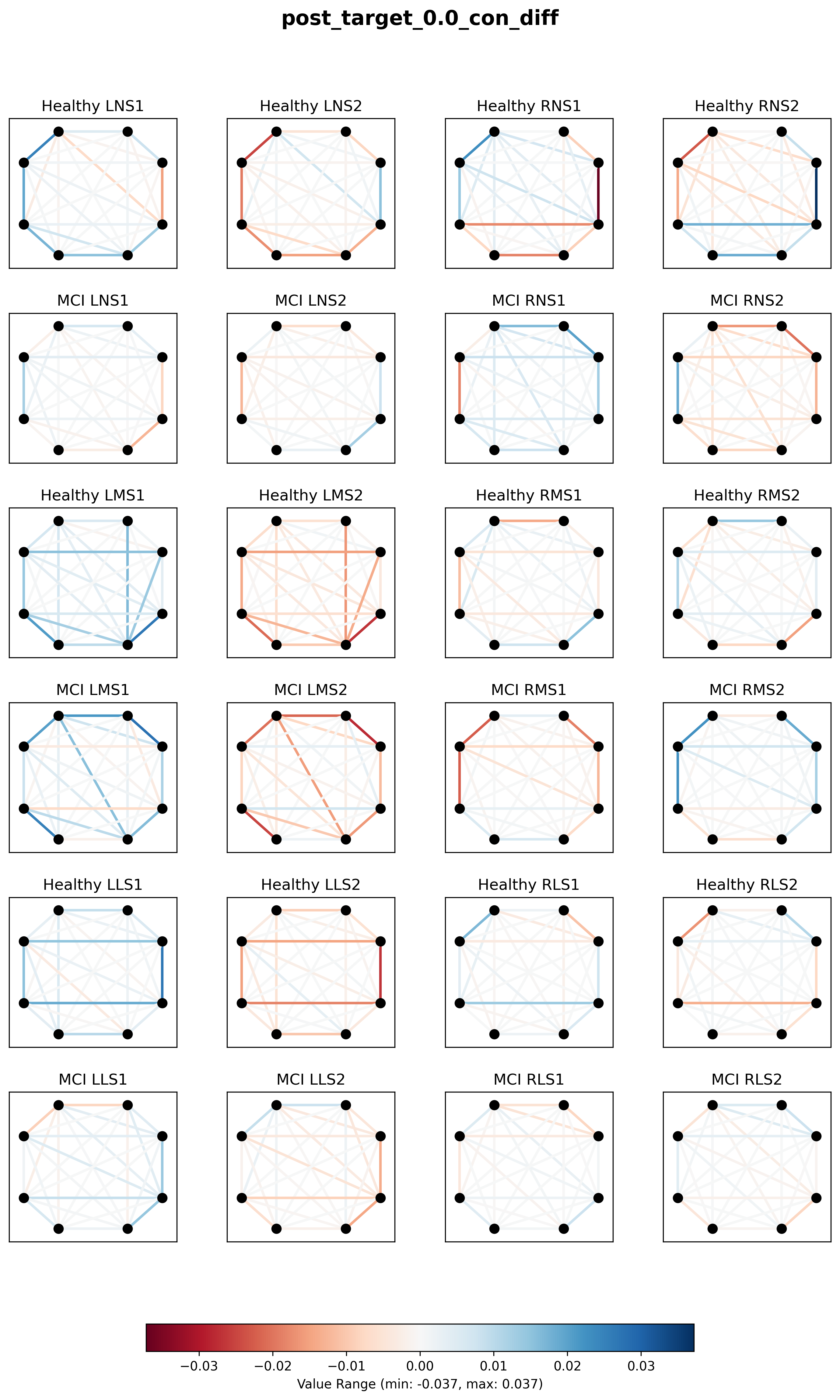}
    \caption{Edge‑wise \textit{Congruent – Incongruent} Routing Efficiency
    (RE) in the 0–0.5 s post‑target window.  
    Each mini‑graph represents one Group $\times$ Saliency $\times$ Target‑Side
    condition (LNS / RNS = left/right non‑salient, LMS / RMS = middle‑salient,
    LLS / RLS = lateral‑salient).  
    Warm (blue) edges indicate higher efficiency on congruent trials; cool
    (red) edges indicate the reverse.}
    \label{fig:RE_con_diff_0to0p5}
\end{figure}

\paragraph{Congruent–incongruent contrast.}
Figure \ref{fig:RE_con_diff_0to0p5} displays edge‑wise
\textit{Congruent – Incongruent} RE maps for every
Group $\times$ Saliency $\times$ Target‑Side combination.  
Healthy controls (top panel in each pair) show
\emph{higher} RE on congruent trials (blue edges) along ipsilateral
prefrontal–frontal and frontal–parietal links, whereas MCI participants
(bottom panels) exhibit the opposite or no clear modulation.
This pattern suggests that healthy adults recruit
right‑hemisphere “short‑cuts” when conflict is low, a mechanism that
appears blunted in MCI.

\begin{figure}[h!]
    \centering
    \includegraphics[width=0.55\textwidth]{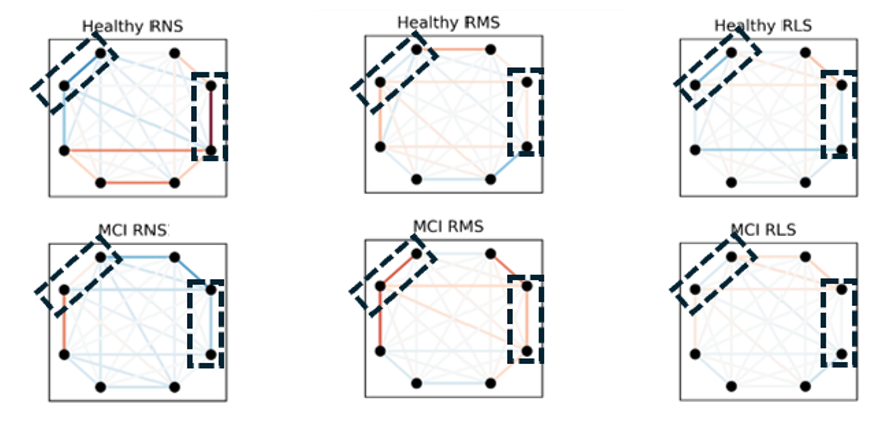}
    \caption{Routing Efficiency difference
    (\textit{Congruent – Incongruent}) for right‑target trials in the
    0–0.5 s window.  
    \textbf{Dashed boxes} highlight the PF‑L $\leftrightarrow$ F‑L and F‑R $\leftrightarrow$ P‑R edges
    that reverse direction between groups: HCs show higher efficiency on
    congruent trials (blue), whereas MCIs do not (red or neutral).  The
    effect is clearest in the Non‑Salient (RNS) column.}
    \label{fig:RE_zoom_0to0p5}
\end{figure}

\paragraph{Right‑target focus.}
Figure \ref{fig:RE_zoom_0to0p5} zooms in on \textbf{right‑target} trials
across the three saliency levels.  
Two fronto-parietal 'shortcut' edges dominate the pattern:

\begin{itemize}
    \item \textbf{PF‑L $\leftrightarrow$ F‑L} (left prefrontal – frontal)
    \item \textbf{F‑R $\leftrightarrow$ P‑R} (right frontal – parietal)
\end{itemize}

For \textbf{Healthy Controls (HC)}, both edges turn \emph{blue},
indicating \emph{higher} Routing efficiency on \emph{congruent}
relative to incongruent trials.  
By contrast, \textbf{MCI} participants show a red shift or no
modulation, implying that these fast frontoparietal 'short cuts' are not recruited - or even down-regulated - when the conflict is minimal.  
The flip in direction is most pronounced in the
\textbf{Non‑Salient / Right‑Target (RNS)} condition, reinforcing the idea
that \emph{MCI fails to engage lateralized fronto‑parietal routes when
they would normally speed processing}.

These preliminary observations suggest that, beyond the loss of global
integration captured by GE, MCI exhibits a pathway‑specific deficit in
fronto‑parietal routing precisely when efficient information transfer would
facilitate rapid and accurate responses.  
Larger samples and formal edge‑wise permutation tests will be required to
determine whether RE can serve as a reliable biomarker of early cognitive
decline.

\section{Conclusion}

By combining whole‑brain \textit{Global Efficiency} with behavioural
indices of \textit{Reaction Time} and \textit{Hit Rate}, we capture a
multidimensional portrait of attentional control in healthy ageing and
mild cognitive impairment (MCI).  
Higher GE, faster RTs and greater accuracy on congruent trials in
Healthy Controls point to a flexible, well‑integrated network that
rapidly reallocates resources when cognitive load is low.  
MCI, in contrast, shows a global loss of integration that manifests as
slower, less accurate performance—evidence that impaired network
efficiency scales directly to everyday cognitive speed and precision.

Our exploratory \textit{Routing Efficiency} analysis extends this picture
from global to pathway‑specific: healthy adults transiently up‑regulate
right‑hemisphere fronto‑parietal “short‑cuts” when conflict is minimal,
whereas MCI does not.  
This focal routing deficit may underlie the behavioural slowing
seen in the flanker task and highlights a potential subnetwork biomarker
for early cognitive decline.

Taken together, the results emphasise that both global and pathway‑level
network measures, linked to behavioural output, are critical for
understanding—and ultimately mitigating—the progression from normal ageing
to MCI.  Such a multiscale perspective can guide the development of
targeted cognitive or neuromodulatory interventions aimed at preserving
network integration and the rapid, accurate information processing it
supports.

\bibliographystyle{apalike}

\end{document}